# Hallucination by Code Generation LLMs: Taxonomy, Benchmarks, Mitigation, and Challenges

YUNSEO LEE*, UNIST, Republic of Korea

JOHN YOUNGEUN SONG*, Handong Global University, Republic of Korea

DONGSUN KIM, Korea University, Republic of Korea

JINDAE KIM, Seoul National University of Science and Technology, Republic of Korea

MIJUNG KIM, UNIST, Republic of Korea

JAECHANG NAM†, Handong Global University, Republic of Korea

Recent technical breakthroughs in large language models (LLMs) have enabled them to fluently generate source code. Software developers often leverage both general-purpose and code-specialized LLMs to revise existing code or even generate a whole function from scratch. These capabilities are also beneficial in no-code or low-code contexts, in which one can write programs without a technical background. However, due to their internal design, LLMs are prone to generating hallucinations, which are incorrect, nonsensical, and not justifiable information but difficult to identify its presence. This problem also occurs when generating source code. Once hallucinated code is produced, it is often challenging for users to identify and fix it, especially when such hallucinations can be identified under specific execution paths. As a result, the hallucinated code may remain unnoticed within the codebase. This survey investigates recent studies and techniques relevant to hallucinations generated by CodeLLMs. We categorize the types of hallucinations in the code generated by CodeLLMs, review existing benchmarks and mitigation strategies, and identify open challenges. Based on these findings, this survey outlines further research directions in the detection and removal of hallucinations produced by CodeLLMs.

## 1 Introduction

Ensuring the accuracy, reliability, and security of code generated by Large Language Models (LLMs) remains a critical challenge [1, 12, 53]. A primary reason for this is the prevalence of hallucinations — instances where the model generates code that is illogical, incorrect, or unfaithful to the specified requirements [14]. Addressing these hallucinations is essential, as they undermine the trustworthiness of the generated code and can introduce significant risks and errors into software applications.

Although benchmarks such as HumanEval [9] and Mostly Basic Python Programming (MBPP) [6] are commonly used to evaluate the code generation performance of LLMs, there remains a lack of standardized methods of assessing the hallucinations generated by CodeLLMs. These general benchmarks only measure the syntactical or token-wise differences between the generated and oracle code. At most, the benchmarks provide simple test cases in which the users can verify a subset of dynamic behaviors of the generated code, which are not useful for defining, detecting, and mitigating hallucinations.

To address hallucination issues of code generation tasks, many researchers have created evaluation benchmarks for the tasks recently, and proposed various approaches to addressing the issues. For example, benchmarks such as

---

*Both authors contributed equally to this research. Yunseo Lee conducted this study while he was an undergraduate student at Handong Global University.
†Corresponding author.

Authors' Contact Information: Yunseo Lee, yunseo.lee@unist.ac.kr, UNIST, Ulsan, Republic of Korea; John Youngeun Song, john.song@handong.edu, Handong Global University, Pohang, Republic of Korea; Dongsun Kim, Korea University, Seoul, Republic of Korea, darkrsw@korea.ac.kr; Jindae Kim, Seoul National University of Science and Technology, Seoul, Republic of Korea, jindae.kim@seoultech.ac.kr; Mijung Kim, UNIST, Ulsan, Republic of Korea, mijungk@unist.ac.kr; Jaechang Nam, Handong Global University, Pohang, Republic of Korea, jcnam@handong.edu.

CodeHaluEval [53] and CodeMirage [1] have been developed to measure hallucination frequencies, while mitigation strategies such as iterative grounding [12] and self-revision feedback loops [37] aim to reduce specific hallucinations.

The goal of this study is to provide a comprehensive analysis of code hallucinations, including their categorization, evaluation metrics, and mitigation strategies. To achieve this goal, we (1) structured a detailed taxonomy of code hallucinations, (2) review and categorize existing benchmarks and evaluation metrics used for detecting these hallucinations, (3) consolidated a list of root causes that contribute to code hallucinations, and (4) survey current mitigation strategies designed to address code hallucinations.

## 2 Differences from other surveys on hallucinations of CodeLLMs

Although hallucinations generated by LLMs in general are studied in multiple surveys [14, 19, 61], our survey focuses on hallucinations observed during code generation tasks using LLMs. The followings are the key aspects of our survey:

- **Focus and Scope**: We focus on hallucinations specifically observed from code generation tasks, addressing unique challenges such as syntactic and semantic discrepancies in code output. In addition, while existing surveys [14, 18, 22, 64] on code generation analyzed performance, benchmarks, data curation, and evaluation metrics, they failed to systematically explore code hallucinations. By exploring taxonomy, benchmarks, metrics, and mitigation strategies tailored to code-specific hallucinations, our survey fills this critical gap and provides a comprehensive framework for future research.
- **Taxonomy and Categorization**: Existing hallucination surveys classify hallucinations into input-conflicting, context-conflicting, and fact-conflicting types [19]. Building upon these classifications, our study introduces a taxonomy that incorporates specialized hallucination types unique to the code generation process, allowing a systematic exploration of hallucination issues specific to this domain.
- **Integration of Benchmarks**: Although other surveys [14, 22, 64] include benchmarks such as HumanEval [9] and TruthfulQA [33], we identified four datasets and benchmarks explicitly aligned with detecting and mitigating code hallucination, such as tests for functional correctness and adherence to APIs.
- **Exploration of Mitigation Strategies**: While previous surveys navigated mitigation approaches for general natural languages [61], we delve into mitigation strategies such as fine-tuning with code-specific datasets, leveraging automated testing frameworks, and integrating static and dynamic program analysis tools for real-time hallucination detection.

## 3 Paper Collection and Review Schema

### 3.1 Survey Scope

We aim to cover in full the taxonomy, benchmarks and evaluation metrics, causes of hallucinations, and mitigation techniques for hallucinations in code generated by CodeLLMs. The criteria for selecting papers are as follows:

- Papers that discuss both LLM-based code generation and LLM hallucination.
- Papers that define code hallucinations or propose taxonomies related to them.
- Papers that propose techniques for detecting or mitigating code hallucinations.
- Papers that introduce datasets or benchmarks for evaluating the performance of CodeLLMs.

To distinguish our study from existing surveys on hallucinations in the Natural Language Processing (NLP) domain and focus on code generation, we included only papers that addressed both *LLM code generation* and *LLM hallucination*. In particular, we searched for papers that explicitly used terms such as *code hallucination* or *hallucinated code*. For mitigation-related studies, we included papers that addressed the correctness of generated code, even if the term *hallucination* was not explicitly mentioned.

## 3.2 Methodology for Literature Identification

We conducted a systematic literature review on various papers. To gather as many relevant studies as possible, Google Scholar keyword searches were performed using the terms "hallucination" and "code generation". Considering the rapid advances in research related to LLMs, the review focused mainly on articles published after 2023, while also including two notable articles from 2022 based on their significance. Titles, abstracts, and introductions of the retrieved papers were manually reviewed and categorized into three main categories: Taxonomy, Benchmark, and Mitigation.

In addition, to ensure comprehensive coverage of studies on code hallucination, the snowball method [59] was employed. Snowballing, commonly used in survey studies, involves tracking citations of identified papers until no additional relevant papers are found. This process helped identify missing studies from the initial search, as well as NLP hallucination papers frequently cited in code hallucination research. Although these NLP studies were not included in the systematic review as they did not focus on code, they provided foundational insights to develop classification criteria for code hallucinations.

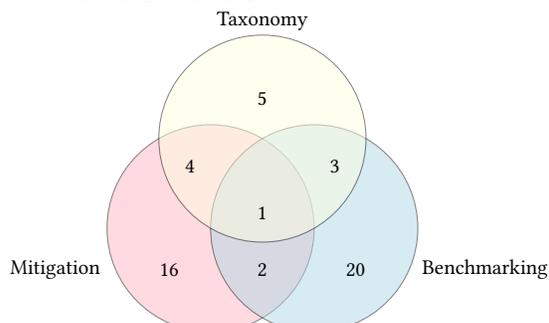

Fig. 1. Distribution of the categorization of papers.

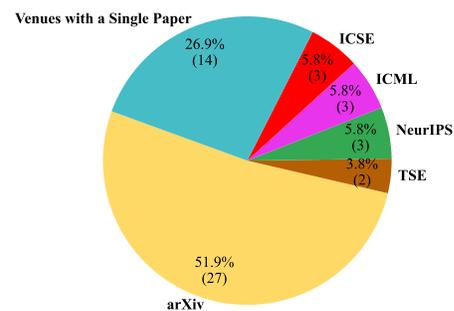

Fig. 2. Distribution of papers by venue.

We categorized the papers into three key dimensions: Taxonomy, Benchmarking, and Mitigation, as shown Fig. 1. Most of the papers fall under the Benchmarking category (20 papers [3, 6–9, 11, 17, 23, 25, 27–30, 35, 38, 49, 50, 66, 66, 67]) and the Mitigation category (16 papers [12, 13, 21, 26, 32, 36, 39, 40, 43, 45, 48, 51, 54, 55, 62, 63]), while fewer studies are categorized under Taxonomy (five papers [15, 24, 42, 52, 57]). Overlapping areas reveal cross-disciplinary contributions: four papers address both Taxonomy and Mitigation [31, 37, 44, 65], three papers address both Taxonomy and Benchmarking [1, 34, 53], and 2 papers explore both Mitigation and Benchmarking [2, 20]. Only one paper [10] combines all three dimensions, emphasizing the scarcity of comprehensive studies.

While many papers are in a preprint stage (e.g., arXiv), authors gradually publish papers at top venues in the community. Fig. 2 shows the distributions of papers by venue. About a half of the papers (51.9%) were published on arXiv. The remaining papers were published in top-tier conferences (39.2%) such as NeurIPS (Annual Conference on Neural Information Processing Systems) and ICML (International Conference on Machine Learning), and academic journals (7.8%) such as TSE (IEEE Transactions on Software Engineering).

## 4 LLM-based code generation (CodeLLMs) and its hallucination

CodeLLMs have been developed to address unique challenges in this domain. OpenAI's Codex and its derivative Copilot are prominent examples that introduced generative pre-trained models with billions of parameters that produce snippets [9, 38]. Following these innovations, models such as Anthropic's Claude Sonnet [5], Meta's CodeLLaMA [46], DeepMind's AlphaCode [30], Salesforce's CodeGen [41], and Amazon's CodeWhisperer [4] entered the landscape, each addressing different aspects of coding efficiency and applicability. OpenAI further refined its offerings with GPT-3.5 and GPT-4, showcasing enhanced capabilities in generating syntactically and semantically accurate code. These advancements are often accompanied by benchmark datasets such as HumanEval [9], DS-1000 [25], and MBPP [6], which assess the performance of LLMs on diverse coding tasks.

Despite their promise, LLMs face a significant challenge in code generation including hallucinations. Hallucinations, in this context, refer to the generation of code that is nonsensical, logically flawed, or unfaithful to the given task description [10]. Studies in the NLP field have classified hallucinations into types such as input-conflicting, context-conflicting, and fact-conflicting hallucinations [19]. Within code generation, hallucinations can manifest as bugs, syntactical errors, security vulnerabilities, or even non-deterministic outputs.Existing research highlights that hallucinated outputs not only degrade functional correctness, but may also introduce subtle errors, such as memory leaks or insecure code [7].

## 5    Taxonomy of Hallucination by CodeLLMs

In our effort to create a consolidated taxonomy of code hallucinations generated by CodeLLMs, we analyzed relevant papers that presented their own classification of hallucinations. Rather than focusing on the causes of the hallucination, our resulting taxonomy categorizes hallucinations based on the observable characteristics of error produced, as shown in Fig. 3. A key advantage of this approach is that it provides an objective for classifying hallucinations, regardless of the model architecture or the training datasets. The taxonomy consists of four primary categories: Syntactic Hallucinations, Runtime Execution Hallucinations, Functional Correctness Hallucinations, and Code Quality Hallucinations. In this section, we discuss each primary category with detailed sub-categories.

### 5.1    Syntactic Hallucinations

These refer to errors that deviate from a language syntax, which render the code unable to parsed and unable to be compiled or interpreted [2, 10, 15, 52, 57]. Syntactic hallucinations can be further classified into two sub-categories: "Syntax Violations" and "Incomplete Code Generation".

*5.1.1    Syntax Violations.* These occur when a CodeLLM generates code that violates the syntax of the programming language, leading to compile-time errors [1, 10, 57]. Three research papers include a specific taxonomy on what kinds of syntax violations there are [1, 10, 57]. One paper [1] classifies errors in generated code that are related to syntax under the term Syntactic Incorrectness. Two papers classify syntax violations further and provide more specific terms such as Incorrect Indentation, Conditional Error, Loop Error, Return Error, and Assignment Error [10, 57].

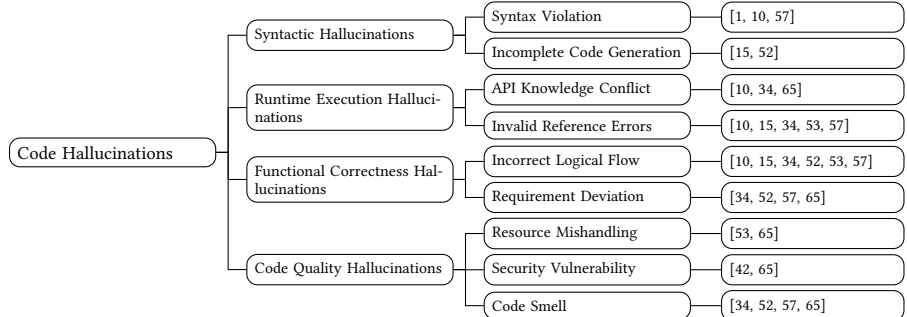

Fig. 3. Taxonomy of hallucinations possibly generated by CodeLLMs.

*5.1.2 Incomplete Code Generation.* This occurs when CodeLLMs stops generating code or entire code blocks are missing [15, 52]. In violation of any specific coding language syntax rules, incomplete code generation will prevent the code from being executed or compiled.

## 5.2 Runtime Execution Hallucinations

These occur when CodeLLMs generate code that is syntactically valid but produces runtime errors, such as exceptions or crashes, during execution [10, 15, 34, 52, 53, 57, 65]. Although syntactic correctness is a necessary condition for code execution, it does not guarantee that the code will function as intended or even run without errors. They manifest only when the code is actually run and may depend on specific inputs or outside factors. Unlike syntactic hallucinations, these types of hallucinations do not necessarily break the syntax, but cause the program to crash or behave unexpectedly.

*5.2.1 API Knowledge Conflict.* This occurs when CodeLLMs misuse libraries or APIs, leading to issues such as missing imports or incorrect or extra parameters [10, 34, 65].

*5.2.2 Invalid Reference Errors.* These arise when CodeLLMs produce code that attempts to access or manipulate program elements that are not yet defined in the code [10, 15, 34, 53, 57]. This can manifest in using variables that have not been declared or attempting to access non-existent members of an object.

## 5.3 Functional Correctness Hallucinations

These arise when CodeLLMs generate code that can execute ,but does not satisfy the functional requirements of the program, which are further categorized as Incorrect Logic Flow and Requirement Deviation [10, 15, 34, 52, 53, 57, 65]. While a program can be syntactically correct and free from runtime errors, it does not guarantee that the code can perform its intended task.

*5.3.1 Incorrect Logical Flow.* This arises when CodeLLMs generates code that contains flaws in their implementation of algorithms and reasoning [10, 15, 34, 52, 53, 57]. These hallucinations often lead to an incorrect solution. This category encompasses flaws such as missing corner cases, incorrect conditional statements, and incorrect arithmetic operations.

*5.3.2 Requirement Deviation.* These arise when CodeLLMs produce code that deviates from the explicit requirements and functionalities outlined in the prompt or problem description [34, 52, 57, 65]. These hallucinations represent the failure of generated code that does not satisfy the requirements of the prompt. Given the diverse situations in which requirement deviation occurs, taxonomies often categorize these errors under broad terms. This category encompasses taxonomy like overall semantic conflicting hallucinations [34] and functional requirement violations [65], while one paper [57] mentions usage of an incorrect function that does not match the requirements.

## 5.4 Code Quality Hallucinations

These occur when CodeLLMs generate code that introduce risks related to resource management, security vulnerabilities, or performance degradation [34, 42, 52, 53, 57, 65]. These hallucinations often compromise the stability, security, and efficiency of the overall system. We categorize these issues into three distinct subcategories: Resource Mishandling, Security Vulnerability, and Code Smell Issues.

Table 1. Comparative Analysis of Code Hallucination Benchmarks.

| Benchmark Name | Language | Number of Tasks | Data Reference | Content | Purpose | Construction |
|---|---|---|---|---|---|---|
| CodeHaluEval [53] | Python | 699 | APPS | Not mentioned | Comparing various types and frequencies of hallucinations in code generation across different LLMs. | Generated code using the APPS dataset, and applied the CodeHalu algorithm to identify the types of hallucinations present and their respective frequencies. |
| CodeMirage [1] | Python | 1,137 | HumanEval, MBPP | Problems, hallucinated code snippets, ground truth code snippets, test cases | Experiment and measure LLM capabilities for automatically detecting code hallucinations using one-shot prompts. | Designed explicit prompts for each of the hallucination types and input them into GPT-3.5 to get Python code generations that have specific hallucination types. |
| LMDefects [15] | Java | 113 (easy: 60, medium: 53) | LeetCode | Problem descriptions, code snippets, public test cases | Evaluate the precision of Codex-generated code and assess the feasibility of applying automated program repair (APR) techniques. | Collected public datasets from LeetCode not included in Codex training. Included a diverse range of Java tasks for analysis. |
| EvalPlus [35] | Python | 164 | HumanEval | Programming tasks, function signatures, and docstrings | Reveal the real correctness of LLM-synthesized code. | Extended the HumanEval dataset by adding type-aware mutations and generating an average of 764.17 test cases per problem to evaluate hallucinations. |
| CodeContests [30] | C++, Java, Python, etc. | 13,328 (training), 117 (validation), 165 (test) | Codeforces, CodeChef, etc. | Problems, Correct and incorrect human submissions, test cases. | To train, validate, and evaluate Alpha-Code. | Leveraged private and public code competition problems. Test cases were expanded through mutation methods. |
| MultiPL-E [8] | 18 languages | Similar to HumanEval, MBPP | HumanEval, MBPP | Not mentioned | Propose the first massively parallel, multi-language benchmark for code generation. | Created a multi-language benchmark by converting Python-based NL2Code benchmarks into 18 programming languages. |
| HalluCode [34] | Python | 5,663 | CodeAlpaca | Objectives, Hallucination categories, Task descriptions | Evaluate the performance of codeLLMs in recognizing hallucinations. | Focused on task description evaluation and detecting hallucinations specific to programming contexts. |

*5.4.1 Resource Mishandling.* These errors arise when CodeLLMs produce code that improperly manages a system's resources, leading to excessive consumption or inefficient allocation of memory that can eventually lead to code failure [53, 65]. Hallucinations like these occur when CodeLLMs write code that includes data processing operations that cause failures due to exceeded memory capacity or when there is numerical overflow due to errors in numerical calculation limits. [53]. Also, Zhang et al. [65] mentions non-functional requirements that are related to suboptimal performance like inefficient loop structures.

*5.4.2 Security Vulnerability.* This arises when CodeLLMs produce code that introduces security weaknesses that make the system susceptible to attacks or unauthorized access [42, 65]. While only two papers have taxonomy that can be categorized under security vulnerabilities, Pearce et al. [42] gives a deep detailed analysis of various security vulnerabilities in generated code. While there are many kinds of security vulnerabilities, some are improper input validation, use after free errors, and null pointer de-reference errors.

*5.4.3 Code Smell.* These occur when CodeLLMs produce code with low maintainability due to extraneous or unnecessary code [34, 52, 57, 65]. Although these hallucinations are not critical for security or performance issues, their absence is crucial for the maintainability and readability of the code that human developers use. These issues include things like dead code, garbage code, or incomplete generation [34, 52, 57]. Sometimes these issues are called "non-functional requirement violation" as code with these issues often contain a part that is unreachable, performs useless assignments, only contains comments, or has empty function bodies [65].

## 6 Benchmarks and Metrics to Evaluate Hallucinations by CodeLLMs

### 6.1 Benchmarks

The growing interest in addressing hallucinations in LLM-generated code has led to the development of various benchmarks. Standard benchmarks are necessary to analyze the hallucination tendencies of various CodeLLMs and to evaluate hallucination detection and mitigation techniques. Table 1 shows recent benchmarks related to code hallucination and summarize their distinct features. Existing benchmarks to evaluate hallucinations by CodeLLMs have limitations, such as a lack of language diversity and a failure to reflect real-world workloads.

Many of those benchmarks build on existing LLM code generation benchmarks, extending them to overcome those limitations. EvalPlus proposed by Liu et al.[35] is a benchmark that extends an existing benchmark, HumanEval, to address its specific limitations. The HumanEval benchmark contains vague task descriptions and insufficient number of test cases per task. Furthermore, some solutions labeled as correct in HumanEval were found to be erroneous. EvalPlus addresses these limitations by increasing the average number of test cases per task to 764.1, leveraging LLMs for seed input generation and employing type-aware mutation for fuzz testing. CodeMirage [1] assesses the ability of LLMs to detect hallucinations in the input code. CodeMirage was generated using the HumanEval and MBPP databases, with artificial hallucinations inserted into the code using the ChatGPT-3.5 model.

Among the seven benchmarks we inspected, five support only one programming language, and four of them (CodeHaluEval, CodeMirage, EvalPlus, and HalluCode) specifically target Python coding tasks. This distribution reflects the frequent use of Python in scenarios where LLMs generate code. In contrast, Fan et al. [15] proposes LMDefects, a Java- focused benchmark that evaluates the correctness of code generated by Codex and explores the applicability of automated program repair (APR) techniques to hallucinated code. LMDefects is based on easy and medium-level problems from the LeetCode platform and incorporates public test cases provided by the platform.

Unlike aforementioned benchmarks, Multiple-E and CodeContests contain code generation tasks in diverse programming languages. Cassano et al. [8] introduced MultiPL-E, a benchmark that translates Python problems from the HumanEval and MBPP datasets into 18 different programming languages. To rigorously compare models, it is essential to evaluate their ability to generate code in languages beyond Python. Multi-language benchmarks have been developed for this purpose, as CodeLLMs are typically designed to handle multiple programming languages. This benchmark uses 18 custom compilers to translate code snippets, test cases, and other components originally designed for Python into other languages, allowing a comparative analysis of LLM performance across languages. These compilers are also extendable to support additional languages in the future.

CodeContests, proposed by Li et al. [30], includes programming challenges from platforms such as Codeforces and CodeChef to train, validate, and evaluate the Alphacode model. This dataset supports multiple programming languages such as C++, Java, Python, etc. enabling broader applicability.

## 6.2 Metrics

To compare and analyze model performance on benchmark datasets that are in line with their research goals, studies adopt different evaluation metrics. Selecting the appropriate metrics is essential to accurately assess the specific aspects of the model that the study aims to target. This section examines evaluation metrics used in the papers that are addressed in Section 6.1. Table 2 summarizes the metrics used in various studies to compare the performance of models with respect to code hallucination. We have grouped the metrics on the following basis: Functional Correctness, Hallucination Detection, Hallucination Recognition and Hallucination Mitigation Metrics.

### 6.2.1 Functional Correctness.

This category focuses on evaluating how well the generated code satisfies its intended requirements. The most common metric, Pass@k, measures the frequency with which at least one of the $k$ generated solutions passes all test cases. Pass@10, a popular variation, represents the fraction of tasks in which at least one of the 10 generated solutions is correct. On the other hand, 10@k measures the percentage of tasks for which $k$ samples were created per task, and when at least 10 of them passed the test. Pass@k and 10@k consider hallucinations in generated code to be any error that prevents the generated code from passing all test cases.

Table 2. Comparative Analysis of Code Hallucination Metrics.

| Category | Metric | Description | Ref. |
|---|---|---|---|
| Functional Correctness | Pass@k | Evaluates the correctness of code generated by a CodeLLM. It measures the likelihood that a CodeLLM generates functionally correct code for a given task. | [8] [15] [35] |
| | 10@k | Evaluates a CodeLLM's ability to generate correct code, specifically to assess the ability to produce multiple correct solutions for a single task. | [30] |
| Hallucination Detection | Hallucination Rate (HR) | Reflects the hallucination phenomenon in LLMs during code generation tasks through actual execution tests. | [53] |
| | Valid Rate (VR) | Reflects the percentage of valid code outputs by an LLM. | [34] |
| | Accuracy of Hallucination Existence Recognition $ACC_{rec}$ | Reflects the percentage of correctly identified existence of hallucinations. | [34] |
| Hallucination Classification | Accuracy of Hallucination Type Recognition $ACC_{type(i)}$ | Reflects the percentage of accurately identified hallucination types. Liu et al. proposed five types of hallucinations. | [34] |
| | Accuracy, Macro-precision, Macro-recall, and Macro-F1 | Standard metrics used to evaluate multi-class classification performance, where classes represent different hallucination types. | [1] |
| Hallucination Mitigation | Accuracy of Hallucination Mitigation $ACC_{mit}$ | Reflects the percentage of modified hallucinated codes which are semantically correct. | [34] |

*6.2.2 Hallucination Detection.* This category quantifies the presence of hallucinations within the generated code. We use Hallucination Rate (HR), Validate Rate (VR) and Accuracy of Hallucination Existence Recognition ($ACC_{rec}$) for this type of metric [34, 53]. HR, as proposed by Tian et al. [53], measures the proportion of generated code samples that syntactically valid but fail to execute as expected using their CodeHalu Algorithm. VR serves as a measure of the proportion of generated outputs that are syntactically valid and executable [34]. Thus, a lower VR can suggest hallucinations are interfering with the code's ability to run. $ACC_{rec}$ used in tandem with VR focuses on how accurately a model identifies valid code outputs that also contain hallucinations.

*6.2.3 Hallucination Type Classification.* This category assesses a CodeLLMs ability to recognize and classify hallucinations. In contrast to detection, type classification aims to categorize given hallucinated codes into one of the predefined hallucination types. The metrics used are Accuracy of Hallucination Type Recognition $ACC_{type(i)}$ [34] and traditional multi-class classification metrics [1]. $ACC_{type(i)}$ assesses the precision of the model in categorizing the type of hallucination present in valid code. Agarwal et al. [1] used accuracy, macro-precision, macro-recall and macro-F1 as metrics to measure how well hallucinations were detected and classified according to their hallucination types. In this context, accuracy refers to the percentage of hallucinations that are well categorized by the model that matches the actual categories.

*6.2.4 Hallucination Mitigation.* This category is used to measure the ability to successfully fix hallucinated codes. Accuracy of Hallucination Mitigation $ACC_{mit}$ [34] shows the percentage of recognized hallucinations that are successfully alleviated by CodeLLMs.

# 7 Causes of Hallucinations in Code Generation

We investigate the causes of hallucinations by CodeLLMs and classify them into three main issues: Training Data Issues, Trained Model Issues, and Prompt Issues. Fig. 4 presents a hierarchical cause analysis tree for code hallucinations generated by CodeLLMs showing the primary causes and into more specific factors.

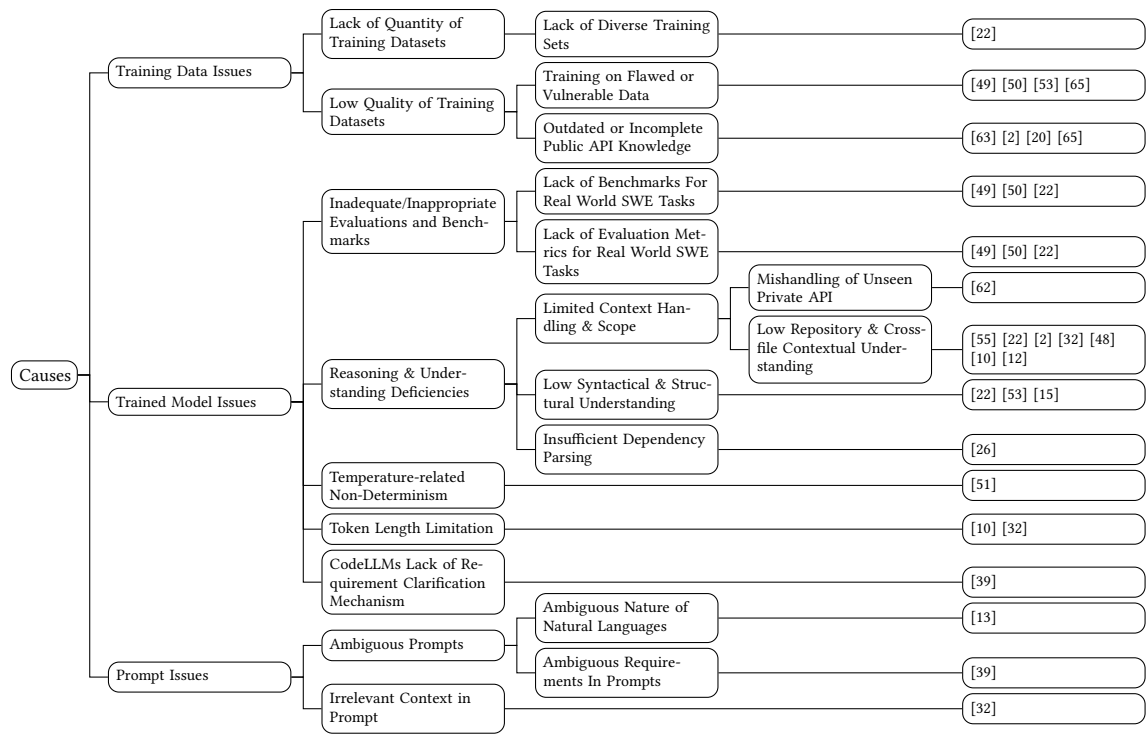

Fig. 4. Potential causes of hallucinations by CodeLLMs.

## 7.1 Training Data Issues

One of the primary causes arises from the issues in quality and quantity of the training data. These can be categorized as follows: a lack of diverse training sets, training on flawed or vulnerable data and, outdated or incomplete public API knowledge. The limited diversity of training data restricts a CodeLLM's ability to generalize across various programming tasks. Jain et al. [20] highlights that the breadth and quality of the training dataset are crucial for correct code generation. In addition, CodeLLMs often produce code hallucinations because of their training on public repositories that often contain deprecated or incomplete API documentation, leading that code to invoke non-existent APIs or contain API misuse [65]. Training on flawed or vulnerable data from open-source projects compound the issues, as these CodeLLMs propagate security vulnerabilities and inefficient implementation to the generated code [50].

## 7.2 Trained Model Issues

Major causes of code hallucinations include issues with the trained models. The causes of code hallucinations that stem from issues with the trained models themselves are: Inadequate or Inappropriate Evaluations and Benchmarks, Reasoning and Understanding Deficiencies, Temperature-related Non-Determinism, Model Input Handling, Token Generation Inefficiency, and CodeLLMs's Lack of Requirement Clarification Mechanism.

One contributor to code hallucinations is the use of inadequate evaluation benchmarks that fail to capture real-world software engineering tasks. Current evaluation metrics and benchmarks often do not accurately represent real-world tasks. CodeLLMs are frequently evaluated using benchmarks that lack the constructs necessary to assess the security of generated code [49, 50]. The need for comprehensive benchmarks and metrics that evaluate a broader spectrum of coding skills is ever prevalent [22].

Another crucial aspect contributing to code hallucinations is the inherent reasoning and understanding deficiencies present in the trained models. One common reasoning and understanding deficiency is the CodeLLM's limited ability to handle code context. As LLMs receive a larger code context, they often mishandle unseen private API and have trouble understanding across files and entire repositories [1, 10, 12, 32, 48, 56]. LLMs lack prior knowledge about private libraries and struggle to leverage external, pre-exisiting knowledge unless they are augmented with retrieval-based generation techniques [62]. This lack of context is evident when generating functions with repository-level dependencies. [56].

The limited capacity of CodeLLMs to grasp the underlying structure and syntax of programming languages highlights their reasoning and understanding deficiencies [15, 22, 53]. As transformer-based LLM architectures are the norm, they may not be optimally designed to fully capture the inherent structure and syntax of programming languages [22]. The CodeLLMs' reliance on pattern matching and statistical rules to generate code, results in a lack of fundamental understanding of symbolic systems [53]. Code is treated as a series of tokens cause language models to lack awareness of program semantics and lead to the generation of incorrect programs [15].

The non-deterministic nature of CodeLLMs, which is controlled by temperature settings and decoding strategies, is an inherent issue with the trained model. The temperature parameter in CodeLLMs governs the randomness of the generated responses as lower temperatures yield more predictable and deterministic outputs, while higher temperatures increase creativity and diversity [51]. While the higher temperatures of verbose models benefit creative code generation, they increase the risk of code hallucination rate [51].

Another aspect contributing to code hallucinations arises from limitations in how trained models handle input tokens. CodeLLMs have an input token length limit, which impacts their ability retain all problem details [10]. This makes it impossible to feed entire code repositories to the CodeLLMs to effectively generate code [32].

The limitations of existing CodeLLMs handling ambiguous requirements can be another source of code hallucinations. Current CodeLLMs often lack a mechanism to clarify unclear or incomplete instructions, which can cause hallucinations that do not satisfy the user's requirements [39].

### 7.3 Prompt Issues

The third major cause of code hallucinations is the prompts. Two factors contributing to this are the ambiguous nature of the prompt itself and presence of insufficient or irrelevant context in the prompt. A significant challenge originates from the inherent ambiguity of natural language prompts. Natural language prompts tend to not fully capture the intent of the user in a fully nuanced and accurate manner. Hence it is a challenge to generate code from such ambiguous natural language prompts [13, 39]. Furthermore, code hallucinations can arise from contextual deficiencies in the prompt. Providing insufficient context or including irrelevant details can hinder the CodeLLM's ability to generate accurate and satisfactory code [32].

## 8 Hallucination Mitigation Methods

Various approaches to mitigate hallucinations are being actively explored. Among these, five approaches were selected for comparative analysis. The following sections provide an overview of the specific hallucination types each approach targets, the root causes they address, and a brief description of each method, along with its strengths and limitations.

### 8.1 De-Hallucinator: Mitigating LLM Hallucinations in Code Generation Tasks via Iterative Grounding

There are two main challenges in addressing this issue. The first challenge is that LLMs lack knowledge of project-specific APIs, and may fail to correctly use existing functions and classes. To investigate this issue, they selected five functions from each of ten open-source projects to create a code completion task. Experimental results showed that 44% of the generated code contained at least one instance of incorrect API usage. To reduce the likelihood of such issues, it would be necessary to provide the entire project code as input. However, due to constraints, this is practically impossible. Therefore, selecting only the essential code snippets to include becomes critical. The second challenge lies in accurately identifying the importance of each piece of code for this purpose. To address this challenge, they suggested an approach named *De-hallucinator* for iteratively retrieving relevant APIs to improve the prompts.

The De-hallucinator [12] pre-analyzes and indexes all source code within the project in advance. When a code generation prompt is provided, it selects the most relevant APIs based on the input and creates a Retrieval Augmented Generation (RAG) prompt to include these APIs. Alternatively, it generates an iterative prompt that incorporates the APIs most relevant to the code produced by the initial prompt. These prompts are then used as inputs for code generation. This approach has the advantage of not requiring modifications to the internal structure of the LLM model. However, it has the drawback of relying on the project to contain well-documented and detailed API descriptions.

### 8.2 Refining ChatGPT-Generated Code: Characterizing and Mitigating Code Quality Issues

Liu et al. [37] proposed a hallucination mitigation method leveraging ChatGPT's self-revision capabilities. The approach aims to address all code quality issues in LLM-generated code, including execution errors, incorrect outputs, and maintainability problems. The method provides two types of feedback to the LLM immediately after code generation: *simple feedback* and *feedback with static analysis*:

- **Simple feedback:** This feedback involves informing the model that the generated code contains quality issues without specifying details.
- **Feedback with static analysis:** This feedback includes more detailed information, such as static analysis results and runtime error messages for the generated code.

The study found that using these feedback methods enabled ChatGPT to self-revise 20–60% of the generated code. Furthermore, iterative feedback led to a gradual improvement in code quality over time.

This approach has the advantage of generalizing scenarios where developers use LLMs for code generation, effectively demonstrating its mitigation performance. However, it has limitations, including the requirement for developers to craft prompts manually and the need for a basic understanding of static analysis tools and error messages.

### 8.3 SynCode: LLM Generation with Grammar Augmentation

Ugare et al. [54] focused on *Syntax Violation Hallucinations*. Grammar-guided generation has recently been widely proposed [16, 43, 47, 58] to ensure that LLM-generated code adheres strictly to predefined grammatical rules [54]. These methods modify the LLM's decoding algorithm to ensure that the model consistently selects tokens conforming to a specific formal language. However, the tokens used by the model are predefined during training, and this often leads to token misalignment where the model's tokens do not match the terminals used in the specified grammar. This misalignment is a significant factor contributing to the high error rates observed in grammar-guided generation. To address this issue, the SynCode algorithm was proposed, leveraging the EBNF (Extended Backus-Naur Form) representation of context-free grammar to guide the LLM during the decoding process. This ensures that the model produces grammatically correct outputs throughout the generation process. The advantage of this approach is its versatility, as it can be applied to any type of LLM decoding algorithm and supports all programming languages.

### 8.4 ClarifyGPT: A Framework for Enhancing LLM-Based Code Generation via Requirements Clarification

Mu et al. [39] proposed a method to mitigate hallucinations caused by ambiguous prompts. Generating correct code requires a clear understanding of the user's requirements, but the necessary information might not always be fully included in the LLM's prompt. In real-world scenarios, developers often address ambiguous requirements by asking clarifying questions to gather additional information. Inspired by this approach, they introduced a novel framework where the LLM generates clarifying questions to help users refine their prompts.

The core challenges of this approach lie in determining when to ask questions and what questions to ask. To address the first challenge, they implemented a code consistency check process. This involves generating test inputs based on the user's prompt and asking the LLM to produce n code solutions aligned with the prompt. The generated code solutions are executed with the test inputs, and the resulting test outputs are compared. If the similarity among outputs is low, it is determined that a clarifying question is needed. This method is based on the intuition that a better understanding of the requirements should result in more consistent code solutions.

For the second challenge, they employed reasoning-based prompts to help the LLM identify elements of the prompt causing ambiguity and generate targeted clarifying questions. The reasoning-based prompt includes instructions for clarifying question generation, few-shot examples, and the user's requirements alongside the generated code solutions.

The ClarifyGPT framework has the advantage of achieving mitigation effects without requiring direct modifications to a model. It also aids developers who struggle to craft clear prompts. However, this approach has significant drawbacks, including high overhead due to the processes of input generation, code generation, and clarifying question generation. Additionally, the examples for the question generation prompt must be manually crafted.

### 8.5 LLM Hallucinations in Practical Code Generation:Phenomena, Mechanism, and Mitigation

Zhang et al. [65] analyzed the types of LLM hallucinations in code generation and potential factors that cause hallucinations. Based on the findings, they suggest a mitigation method based on RAG. The study identified three primary root causes of hallucinations in LLM-generated code: (1) incorrect or insufficient understanding of task requirements, (2) lack of factual knowledge relevant to the generation tasks, and (3) inability to access the necessary code and non-code resources from the repository. To mitigate these issues, the authors proposed a RAG-based approach. They first created a retrieval corpus by scanning all source files from repositories in the CoderEval dataset and extracting consecutive lines of code. When a query is presented to the LLM, the system retrieves related code snippets from the corpus, appending the most relevant ones to the prompt.

This approach has several advantages. It requires no additional effort from users, ensures that only essential information necessary for code generation is provided to the model, and supports handling project-specific APIs. However, its effectiveness is significantly influenced by the quality and quantity of the source code available for retrieval. Moreover, the retrieval process introduces overhead, which can impact efficiency.

Despite these challenges, the RAG-based mitigation method demonstrated a modest reduction in hallucinations across six LLMs. This study serves as a pilot exploration of RAG-based mitigation methods, shedding light on their possible applications in reducing hallucinations in LLMs.

## 9 Discussion and conclusion

The findings in this paper suggest several promising directions for future research. First, the development of more diverse and representative benchmark datasets, encompassing various programming languages and use cases, is essential for evaluating LLMs in broader contexts. Second, advances in hallucination mitigation techniques, such as retrieval-augmented generation, clarifying question frameworks, and grammar-guided decoding, indicate the potential of combining multiple approaches to enhance reliability. Third, the integration of LLMs into real-world software development workflows calls for adaptive techniques that can dynamically address context-specific hallucinations, improving practical usability. By synthesizing these insights, this study serves as a road-map for advancing research and development in LLM code generation, ultimately contributing to the creation of more robust and trustworthy systems.